%% file: 0main.tex
\documentclass[10pt,conference]{IEEEtran}

\usepackage[utf8]{inputenc}
\usepackage{amsmath}
\usepackage{amsfonts}
\usepackage{amsthm}
\usepackage{amssymb}
\usepackage{optidef}
\usepackage{multirow}
\usepackage{ulem}
\usepackage{threeparttable}
\usepackage{array}
\usepackage{algorithm}
\usepackage{algorithmic}
\usepackage{graphicx}
\usepackage{subfigure} 
\usepackage{microtype} 
\usepackage[table,xcdraw]{xcolor}
\usepackage{cite}
\title{Resource Allocation for ISAC Networks with Application to Target Tracking}

\author{
\IEEEauthorblockN{
Lu Wang\IEEEauthorrefmark{1}, 
Luis F. Abanto-Leon\IEEEauthorrefmark{2}} 
\IEEEauthorblockA{\IEEEauthorrefmark{1}Department of Computer Science, Technical University of Darmstadt, Darmstadt, Germany\\}
\IEEEauthorblockA{\IEEEauthorrefmark{2}Faculty of Electrical Engineering and Information Technology, Ruhr University Bochum, Bochum, Germany\\}
Email: \IEEEauthorrefmark{1}lwang@wise.tu-darmstadt.de, \IEEEauthorrefmark{2}luis.abantoleon@ruhr-uni-bochum.de
}

\begin{document}

\maketitle

\input{1Abstract}

\input{2Intro}
\input{3systModel}

\input{4probForm}
\input{5Solution}
\input{6Results}
\input{8Conclusion}


\end{document}

%% file: 1Abstract.tex
\begin{abstract}


Future 6G networks are expected to empower communication systems by integrating sensing capabilities, resulting in integrated sensing and communication (ISAC) systems. However, this integration may exacerbate the data traffic congestion in existing communication systems due to limited resources. Therefore, the resources of ISAC systems must be carefully allocated to ensure high performance. Given the increasing demands for both sensing and communication services, current methods are inadequate for tracking targets frequently in every frame while simultaneously communicating with users. To address this gap, this work formulates an optimization problem that jointly allocates resources in the time, frequency, power, and spatial domains for targets and users, accounting for the movement of targets and time-varying communication channels. Specifically, we minimize the trace of posterior Cramér-Rao bound for target tracking subject to communication throughput and resource allocation constraints. To solve this non-convex problem, we develop a block coordinate descent (BCD) algorithm based on the penalty method, successive convex approximation (SCA), and one-dimensional search. Simulation results demonstrate the validity of the proposed algorithm and the performance trade-off between sensing and communication.

\let\thefootnote\relax\footnotetext{This work was supported by the Deutsche Forschungsgemeinschaft [German Research Foundation (DFG)] mm-Cell project under Grant 520 01825.}

\end{abstract}

\begin{IEEEkeywords}
ISAC, resource allocation, target tracking, dynamicity
\end{IEEEkeywords}

%% file: 2Intro.tex
\section{Introduction}
The advent of mobile communication has completely revolutionized the world since the 1980s, especially 4G and 5G systems. In order to pursue a better life experience, the next-generation communication system, 6G, is nowadays being researched far beyond just communications~\cite{HWwitepaper}. One of the typical usage scenarios for 6G systems, announced by the International Telecommunication Union (ITU) organization, is integrated sensing and communication (ISAC)~\cite{IMT2030}. Compared with traditional communication-only cellular networks, ISAC supports both high-resolution sensing and communication in one unified system, saving hardware costs and energy. Moreover, the sensing integrated 6G network is able to obtain sensory information from the environment, which can be exploited to build ubiquitous intelligence for the network~\cite{FanLOverview}. Despite the advantages of ISAC, the integration of sensing and communication consumes more resources, which may exacerbate the data traffic congestion in existing communication systems. Therefore, the ISAC system needs to be carefully designed to satisfy performance requirements.

To solve the foregoing congestion problem, on the one hand, ISAC systems are evolving to be deployed in millimeter-wave (mmWave) bands with large bandwidths to support high throughput and sensing resolution. On the other hand, the limited resources of ISAC systems need to be carefully allocated so that the integration effectively utilizes the resources~\cite{RAsurvey}. Therefore, in this paper, we focus on the resource allocation of the ISAC system in mmWave bands. To date, resource allocation of ISAC systems has been investigated with different design goals in various scenarios. The works~\cite{KLD, UAV5} allocated transmit power to users and targets in target detection and unmanned aerial vehicle (UAV) scenarios. With more sensing and communication tasks, the works~\cite{MEC4,dong2022sensing, UAV6} involved one more dimension of frequency resource, either allocating bandwidth in~\cite{MEC4} or jointly assigning power and bandwidth in~\cite{dong2022sensing, UAV6}. Furthermore, in a system with multiple users and targets, beamforming is an effective way to efficiently focus the spatial resource on the served users, especially under mmWave bands with stronger path loss. The authors in~\cite{khalili2023energy,ding2022joint} selected beams or designed beamforming to improve sensing and communication performances in ISAC systems.

However, the aforementioned resource allocation schemes simultaneously serving users and targets are designed for one snapshot. When users or targets move in different time slots, the channel state information (CSI) also varies. Correspondingly, the resource allocation needs to be updated in new slots. Therefore, for practicality, it is necessary to consider the resource allocation for multiple time slots instead of just one snapshot. The existing ISAC works~\cite{TVTTckPT, UAV4, wu2022resource} studied the time slot allocation among users, where the access of users and targets is served in a time-division manner. Note that this time-domain scheduling differs from the necessity of reallocating all the resources at different time slots due to target movement or channel variation. In this work, we refer to the phenomenon of user movement or channel variation as dynamicity. The works~\cite{xu2022robust,chen2023impact} designed the systems under dynamicity. In~\cite{xu2022robust}, the duration of multiple snapshots was optimized under the imperfect CSI caused by dynamicity in the scenario of physical layer security. However, the detected target in this context is a potential eavesdropper, whose performance is desired to degrade rather than improve. The authors in~\cite{chen2023impact} considered the effect of the dynamicity on the Cram\'er-Rao bound (CRB) for target tracking and the achievable rate for communication by allocating frequency and power resources and optimizing channel re-estimation interval. However, the target tracking occurs only at the beginning of each frame, which does not meet the increasing sensing requirements of future 6G networks. 

To fill in the foregoing research gap, we design a transmission scheme for the ISAC system under dynamicity where the target tracking and the user equipments (UEs) communication take place at every frame. When dynamicity exists, the CSI estimation needs to be updated frequently, which is impractical. Therefore, we apply an extended Kalman filter (EKF) to predict and track the state of targets so that the instant CSI between the base station (BS) and targets is not needed. Moreover, to efficiently utilize resources, an optimization problem is formulated by minimizing the trace of posterior CRB (PCRB) of targets subject to the communication throughput and practical resource constraints. To solve the formulated non-convex problem, we propose a block coordinate descent (BCD) algorithm based on the penalty method, successive convex approximation (SCA), and one-dimensional search. Finally, the validity of the proposed solution and the trade-off between sensing and communication are analyzed via simulation results.

%% file: 3systModel.tex
\section{System Model}
\begin{figure}[htbp]
	\centering
	\includegraphics[width=1\linewidth]{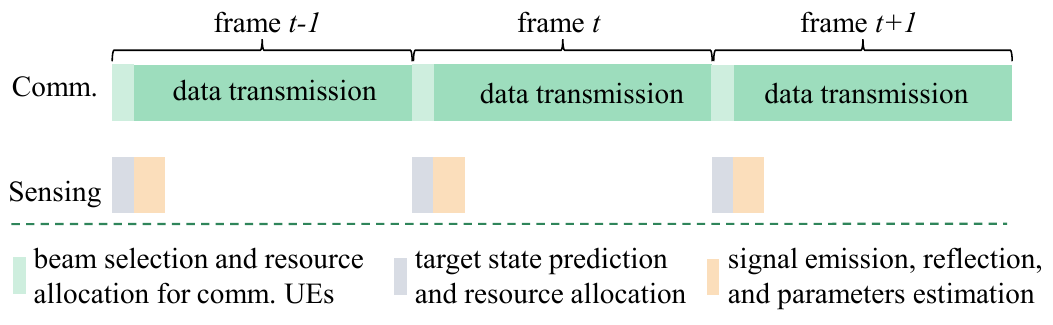}
	\caption{Transmission protocol of ISAC system.}
	\label{TransProto}
\end{figure}

We consider a mmWave ISAC network consisting of $M$ targets, $K$ single-antenna UEs, and a BS  equipped with a uniform linear array (ULA) with $N_t$ transmit antennas and $N_r$ receive antennas, which performs joint communication and location-tracking across multiple frames. Fig.~\ref{TransProto} shows the transmission protocol, where the resources for UEs and targets are allocated at the beginning of each frame. The states of the targets are also predicted. Afterward, the signal transmission and reflection take place. With limited resources available, we aim to allocate resources for targets and UEs to fulfill their respective performance requirements.

\subsection{Resource Allocation Model}
\label{RAmodel}
We consider four-dimensional resource allocation, including resource blocks (RBs) in the frequency domain, mini-slots in the time domain, power in the power domain, and beams in the spatial domain. The mini-slot, which serves as the scheduling unit in the time domain, consists of several orthogonal frequency-division multiplexing (OFDM) symbols~\cite{38912}. There are $N_\mathrm{RB}$ RBs in the frequency domain and $U$ frames in the time domain. Each frame has a duration of 10 ms and consists of 10 sub-frames. We designate one mini-slot to contain 7 OFDM symbols, which can be easily extended to other numbers of symbols. The number of mini-slots per frame, denoted by $I$, varies based on the considered numerology. Thus, the total number of resource units (RUs), each comprising one RB in the frequency domain and one mini-slot in the time domain, in one frame is $I \times N_\mathrm{RB}$.

We allocate contiguous RBs for targets in the frequency domain. To avoid interference, UEs and targets are allocated with orthogonal resources, i.e., communication and sensing occur in disjoint resources~\cite{dong2022sensing}. Since distance measurement performance is determined by bandwidth, we allocate one mini-slot but as many RBs as possible for each target. For each frame $u$, let $n_{m}^{u} \in \{1,\dots, N_\mathrm{RB}-1\}$ denote the number of contiguous RBs allocated to the $m$-th target. Let ${o_{ki}^u} \in \{1,\dots, N_\mathrm{RB}-1\}$ represent the number of RBs allocated to the $k$-th UE in the $i$-th mini-slot. The total number of RUs allocated to UEs and targets in each frame is no larger than the maximum number of RUs, mathematically expressed as
\begin{equation}\label{ContinuRA}
\sum\limits_{m = 1}^{M} {n_m^u}  + \sum\limits_{k = 1}^{K} \sum\limits_{i = 1}^I {o_{ki}^u} \le I \times {N_{RB}},\forall u \in U.
\end{equation}
Meanwhile, Let $B_0$ and $T_0$ denote the bandwidth of one RB and the duration of one mini-slot. Therefore, $n_m^u{B_0}$ represents the allocated bandwidth for the $m$-th target at the $u$-th frame. ${o_{ki}^u}{B_0}$ represents the allocated bandwidth to the $k$-th UE in the $i$-th mini-slot of $u$-th frame.

\subsection{Sensing Model}
\subsubsection{Sensing signal model}
During the $u$-th frame, the reflected sensing signals of $M$ targets received by the BS at time instant $t$ are expressed as follows.
\begin{equation}\label{SenEcho}
{\bf{r}}_s^u\left( t \right) = \sum\limits_{m= 1}^{M} {\sqrt {{p_m^u}} {\bf{H}}_m^u{\bf{f}}_m^us_m^u\left( {t - \tau _m^u} \right){e^{j2\pi v_m^ut}} + {\bf{w}}_s^u\left( t \right)},
\end{equation}
where for the $m$-th target, $s_m^u\left( t \right)$ is the sensing signal. $p_m^u$ refers to the allocated power. ${\bf{H}}_m^{u} = \alpha _m^{u}{{\bf{a}}_r}\left( {\phi _m^{u}} \right){\bf{a}}_t^{\rm{H}}\left( {\theta _m^{u}} \right)$ represents the sensing channel. ${\alpha _m^{u}}$ is the reflection coefficient decided by the array gain $\sqrt {{N_t}{N_r}}$, the carrier wavelength $\lambda$, the radar cross section $\sigma$, and the distance ${d_m^{u}}$ between the target and the BS, mathematically satisfying ${\left( {\alpha _m^{u}} \right)^2} = {{{N_t}{N_r}{\lambda ^2}\sigma } \mathord{\left/ {\vphantom {{{N_t}{N_r}{\lambda ^2}\sigma } {\left( {4{\pi ^3}{{\left( {d_m^{u}} \right)}^4}} \right)}}} \right. \kern-\nulldelimiterspace} {\left( {4{\pi ^3}{{\left( {d_m^{u}} \right)}^4}} \right)}}$. $\tau _m^{u}$ and $v_m^{u}$ are time delay and Doppler frequency shift, respectively. ${{\bf{a}}_t}\left( {\theta _m^{u}} \right)$ is the transmit steering vector with the angle of departure (AoD) at ${\theta _m^{u}}$ given by
\begin{equation}\label{TrSteerVec}
{{\bf{a}}_t}\left( {\theta _m^{u}} \right) = \sqrt {\frac{1}{{{N_t}}}} {[1,{e^{j\frac{{2\pi }}{\lambda }d\sin (\theta _m^{u})}},...,{e^{j\frac{{2\pi }}{\lambda }d({N_t} - 1)\sin (\theta _m^{u})}}]^{\rm{T}}},
\end{equation}
where the antenna space $d$ and the wavelength $\lambda$ satisfy $d = {\lambda  \mathord{\left/{\vphantom {\lambda  2}} \right.\kern-\nulldelimiterspace} 2}$. ${{\bf{a}}_r}\left( {\phi _m^{u}} \right)$ is the receive steering vector with the angle of arrival (AoA) at ${\phi _m^{u}}$, which has the same expression as (\ref{TrSteerVec}) but replacing $N_t$ with $N_r$. ${{\bf{w}}_s^{u}}$ is the complex Gaussian noise with zero mean and variance $\sigma _s^2$ for each element. ${\bf{f}}_m^{u} $ is the beamforming vector, pointing to the predicted AoD ${\hat \theta _m^{u}}$\footnote{The predicted AoD can be obtained based on (\ref{Coordi2Polar}) and (\ref{StatePre}).} of the $m$-th target, namely
\begin{equation}\label{BFsen}
{\bf{f}}_m^{u} = {{\bf{a}}_t}\left( {\hat \theta _m^{u}} \right).
\end{equation}

We assume that all the targets move with constant velocity from frame to frame. Under the Cartesian coordinate system, we suppose the BS is located at $\left( 0,0 \right)$. The $m$-th target has the initial location $\left( {x_m^{0},y_m^{0}} \right)$ and velocity $\left( {\dot x_m^{0},\dot y_m^{0}} \right)$. The state of the $m$-th target at the $u$-th frame is denoted as ${\bf{\xi }}_m^{u} = {\left[ {x_m^{u},y_m^{u},\dot x_m^{u},\dot y_m^{u}} \right]^{\rm{T}}}$, where $\left( {x_m^{u},y_m^{u}} \right)$ and $\left( {\dot x_m^{u},\dot y_m^{u}} \right)$ represent the location and velocity of this target, respectively. To track the target locations, we introduce a state prediction model and a measurement model. Afterwards, the sensing metric PCRB is derived to measure the localization tracking performance.

\subsubsection{State evolution model}
The motion evolution for the $m$-th target between adjacent frames $u$ and $u-1$ is given by~\cite{yan2015simultaneous}.
\begin{equation}\label{TarMotion}
{\mathbf{\xi }}_m^{u} = {{\mathbf{F}}_\xi }{\bf{\xi }}_m^{u-1} + {\bf{w}}_m^{u-1},
\end{equation}
where ${{\mathbf{F}}_\xi }$ is the transition matrix between two frames, namely
\begin{equation}\label{TransiMat}
{{\bf{F}}_\xi } = \left[ {\begin{array}{*{20}{c}}
1&{{T_s}}\\
0&1
\end{array}} \right] \otimes {{\bf{I}}_2},
\end{equation}
$T_s$ is the time interval between adjacent frames. ${{\bf{I}}_2}$ refers to the 2-order identity matrix. $\otimes$ is the Kronecker product. ${\bf{w}}_m^{u-1}$ models the process noise at ($u-1$)-th frame and follows Gaussian distribution with zero mean and covariance ${\bf{\Phi }}_m$, i.e.
\begin{equation}\label{PNCovariMat}
{\bf{\Phi }}_m = \left[ {\begin{array}{*{20}{c}}
{\frac{1}{3}T_s^3}&{\frac{1}{2}T_s^2}\\
{\frac{1}{2}T_s^2}&{{T_s}}
\end{array}} \right] \otimes \tilde \sigma _m{{\bf{I}}_2},
\end{equation}
where $\tilde \sigma _m$ represents the process noise level of the $m$-th target.

\subsubsection{Measurement model}
The targets can be localized by their distance and AoA towards BS, which are measured by 
\begin{equation}\label{MeasureModPosition}
{\bf{z}}_m^u = {\bf{g}}\left( {{\bf{\xi }}_m^u} \right) + {\bf{\tilde w}}_m^u,
\end{equation}
where ${\bf{g}}\left( {{\bf{\xi }}_m^u} \right) = {\left[ {d_m^{u},\phi _m^{u}} \right]^{\rm{T}}}$ collects the distance $d_m^{u}$ and AoA $\phi _m^{u}$, obtained by the mapping function shown as follows.
\begin{equation}\label{Coordi2Polar}
{\bf{g}}\left( {{\bf{\xi }}_m^u} \right) = \left\{ {\begin{array}{*{20}{c}}
{d_m^u = \sqrt {{{\left( {x_m^u} \right)}^2} + {{\left( {y_m^u} \right)}^2}} ,}\\
{\phi _m^u = \arctan \left( {y_m^u/x_m^u} \right).}
\end{array}} \right.
\end{equation}
$\widetilde {\bf{w}}_m^{u}$ is the measurement error following Gaussian distribution with zero mean and covariance matrix ${\bf{\Sigma }}_m^{u}$ given by
\begin{equation}\label{MeErrorCov}
{\bf{\Sigma }}_m^{u} = {\rm{diag}}\left( {\sigma _{d_m^{u}}^2,\sigma _{\phi _m^{u}}^2} \right)
\end{equation}
where $\sigma _{d_m^{u}}^2$ and $\sigma _{\phi _m^{u}}^2$ are the CRBs of distance and AoA estimation of the $m$-th target at the $u$-th frame, given by \cite{yan2015simultaneous}
\begin{equation}\label{CRBd}
\sigma _{d_m^{u}}^2 \propto {\left( {{p_m}\left\| {{\bf{H}}_m^{u}{\bf{f}}_m^{u}} \right\|_2^2{{\left( {n_m^{u}{B_0}} \right)}^2}} \right)^{ - 1}},
\end{equation}
\begin{equation}\label{CRBtheta}
\sigma _{\phi _m^{u}}^2 \propto {\left( {{p_m}\left\| {{\bf{H}}_m^{u}{\bf{f}}_m^{u}} \right\|_2^2/{\varpi _{nn}}} \right)^{ - 1}},
\end{equation}
where ${\varpi _{nn}}$ is the null to null beamwidth of receive antenna.

\subsubsection{Posterior CRB}
We use PCRB to measure the location tracking performance. To obtain PCRB, we first calculate the posterior Fisher information matrix (FIM) ${\bf{J}}\left( {{\bf{\xi }}_m^{u}} \right)$, given by~\cite{dong2022sensing}.
\begin{equation}\label{PFIM}
{\bf{J}}\left( {{\bf{\xi }}_m^u} \right) = {{\bf{J}}_p}\left( {{\bf{\xi }}_m^u} \right) + {{\bf{J}}_d}\left( {{\bf{\xi }}_m^u} \right),
\end{equation}
where ${{\bf{J}}_p}\left( {{\bf{\xi }}_m^{u}} \right)$ denotes the prior FIM and can be recursively obtained using the previous FIM, namely
\begin{equation}\label{FIMP}
{{\bf{J}}_p}\left( {{\bf{\xi }}_m^{u}} \right) = {\left[ {{\bf{\Phi }}_m^{u-1} + {{\bf{F}}_\xi }{{\bf{J}}^{ - 1}}\left( {{\bf{\xi }}_m^{u-1}} \right){{\left( {{{\bf{F}}_\xi }} \right)}^{\rm{T}}}} \right]^{ - 1}}.
\end{equation}
${{\bf{J}}_d}\left( {{\bf{\xi }}_m^{u}} \right)$ is the data FIM and can be approximated as follows.
\begin{equation}\label{FIMD}
{{\bf{J}}_d}\left( {{\bf{\xi }}_m^{u}} \right) = {\left( {{\bf{\hat Q}}_m^{u}} \right)^{\rm{T}}}{\left( {{\bf{\hat \Sigma }}_m^{u}} \right)^{ - 1}}{\bf{\hat Q}}_m^{u},
\end{equation}
where ${{\bf{\hat Q}}_m^{u}}$ means the Jacobian matrix of ${\bf{g}}\left( {{\bf{\xi }}_m^{u}} \right)$ evaluated at the predicted target state ${\bf{\hat \xi }}_m^{u}$. ${{\bf{\hat \Sigma }}_m^{u}}$ is the covariance matrix of the measurement error evaluated at the predicted state ${\bf{\hat \xi }}_m^{u}$, which depends on the allocated resources. Therefore, given the previous FIM at the $(u-1)$-th frame, we can predict the FIM at the $u$-th frame based on the resources allocated to targets. The predictive PCRB is the inverse of the posterior FIM, i.e. $ {{\bf{J}}{{( {{{\bf{\xi }}_m}} )}^{ - 1}}} $. The location tracking performance of each target at each frame is evaluated by the trace of PCRB ${\rm{trace}}{( {{\bf{J}}( {{\bf{\xi }}_m^u} )} ^{ - 1})}$.

\addtolength{\topmargin}{0.02in} 

\subsubsection{Prediction based on EKF}
\label{PreEKL}
After receiving the reflected echos, distance and AoA are estimated. Then, the state of targets at the next frame is predicted using the EKF, which is used to keep track of targets. Thus, we have~\cite{dong2022sensing}

\begin{itemize}
\item Step 1: State Prediction
\begin{equation}\label{StatePre}
{\bf{\hat \xi }}_m^{u|u - 1} = {{\bf{F}}_\xi }{\bf{\hat \xi }}_m^{u - 1}.
\end{equation}
\item Step 2: MSE Matrix Prediction
\begin{equation}\label{MSEpre}
{\bf{M}}_m^{u|u - 1} = {{\bf{F}}_\xi }{\bf{M}}_m^{u - 1}{\bf{F}}_\xi ^{\rm{T}} + {{\bf{\Phi }}_m}.
\end{equation}
\item Step 3: Kalman Gain Calculation
\begin{equation}\label{KalmanGain}
{\bf{K}}_m^u = {\bf{M}}_m^{u|u - 1}{\left( {{\bf{\hat Q}}_m^u} \right)^{\rm{T}}}{\left( {{\bf{\hat \Sigma }}_m^u + {\bf{\hat Q}}_m^u{\bf{M}}_m^{u|u - 1}{{\left( {{\bf{\hat Q}}_m^u} \right)}^{\rm{T}}}} \right)^{ - 1}}.
\end{equation}
\item Step 4: State Tracking
\begin{equation}\label{StateTrack}
{\bf{\hat \xi }}_m^u = {\bf{\hat \xi }}_m^{u|u - 1} + {\bf{K}}_m^u\left( {{\bf{z}}_m^u - {\bf{g}}\left( {{\bf{\hat \xi }}_m^{u|u - 1}} \right)} \right).
\end{equation}
\item Step 5: MSE Matrix Update
\begin{equation}\label{MSEupdate}
{\bf{M}}_m^u = \left( {{\bf{I}}_4 - {\bf{K}}_m^u{\bf{\hat Q}}_m^u} \right){\bf{M}}_m^{u|u - 1}.
\end{equation}
\end{itemize}

\subsection{Communication Model}
The downlink channel from the BS to the $k$-th UE ${\bf{h}}_{ck}^{u}$ varies from frame to frame, modeled as follows.
\begin{equation}\label{HCOMtime}
{\bf{h}}_{ck}^{u} = {\rho _k}{\bf{h}}_{ck}^{u-1} + \sqrt {1 - \rho _k^2} {\bf{\mathord{\buildrel{\lower3pt\hbox{$\scriptscriptstyle\frown$}} 
\over w} }}_k^{u},
\end{equation}
where the frequency selectivity is assumed to not exist while the temporal dynamicity is considered \cite{chen2023impact}. ${\rho _k}$ is the time correlation coefficient between adjacent frames. ${\bf{\mathord{\buildrel{\lower3pt\hbox{$\scriptscriptstyle\frown$}} 
\over w} }}_k^{u} $ represents the complex Gaussian noise with zero mean and variance $\sigma _h^2$ for each element. The channel of the $k$-th UE at the first frame ${\bf{h}}_{ck}^{1}$ is modeled as the geometric channel model under mmWave bands, given by
\begin{equation}\label{HCOMk}
{\bf{h}}_{ck}^{1} = \sqrt {\frac{{{N_t}}}{{{N_P}}}} \sum\limits_{p = 1}^{{N_P}} {\tilde \alpha _{kp}^{u}{\bf{a}}_t^{\rm{H}}\left( {\theta _{kp}^{u}} \right)},
\end{equation}
where $N_P$ is the number of paths. ${\tilde \alpha _{kp}^{u}}$ and ${\theta _{kp}^{u}}$ are the channel coefficient and AoD of the $p$-th path, respectively. 

To deal with the multi-user interference and the path loss under mmWave bands, we employ codebook-based beamforming to reduce the complexity of beamforming, where beams have normalized power. Meanwhile, power allocation is applied to utilize the power more efficiently. The received signals of the $k$-th UE at time instant $t$ are expressed as follows.
\begin{equation}\label{receCOM}
r_{ck}^u\left( t \right) = \sqrt {{p_k^u}} {\bf{h}}_{ck}^u{\bf{f}}_k^{u}{\bf{s}}_k^u\left( {t - \tau _k^u} \right){e^{j2\pi v_k^ut}} + w_c^u\left( t \right),
\end{equation}
where $p_k^u$, ${\bf{f}}_k^{u}$, ${\tau _k^{u}}$ and $v_k^{u}$ are the transmit power, beamformer, time delay, and Doppler shift, respectively. $w_c^{u}$ is the complex Gaussian noise with zero mean and variance $\sigma _c^2 = o_{ki}^u{B_0}{N_0}$ for each element. $N_0$ is the noise power spectrum density. The throughput of each UE at each frame is calculated as follows
\begin{equation}\label{TPk}
R_k^u{(o_{ki}^u,p_k^u,{\bf{f}}_k^{u})} = \sum\limits_{i = 1}^I {(o_{ki}^u{B_0}){{\log }_2}\left( {1 + \frac{{{p_k^u}{{\left| {{\bf{h}}_{ck}^u{\bf{f}}_k^{u}} \right|}^2}}}{{o_{ki}^u{B_0}{N_0}}}} \right)}.
\end{equation}

%% file: 4probForm.tex
\section{Problem Formulation}

To improve both sensing and communication performance, we jointly allocate power, bandwidth, time, and spatial resources to UEs and targets by formulating and solving an optimization problem. We minimize the trace of PCRB of all targets while satisfying the UEs' communication throughput requirements and the practical resource allocation constraints in all $U$ frames, which is mathematically formulated as
\begin{subequations}\label{OptP1}
\begin{align}
({\rm P}1) & \mathop {\min }\limits_{\{ n_m^u,o_{ki}^{u},p_m^u,p_k^u,{\bf{ f}}_k^{u} \} } {\sum\limits_{u = 1}^U {\sum\limits_{m = 1}^{M} {{\rm{trace}}{\left( {{\bf{J}}\left( {{\bf{\xi }}_m^u} \right)} ^{ - 1}\right)}} } } \label{P1Obj}\\
\mbox{s.t.}
&\sum\limits_{i = 1}^I {R_k^u{(o_{ki}^u,p_k^u,{\bf{f}}_k^{u})}}  \ge \gamma _k^u,\forall k \in {\cal K},\forall u \in {\cal U}, \label{const_TP}\\
& \sum\limits_{m = 1}^{M} {p_m^u} + \sum\limits_{k = 1}^{K} {p_k^u} \le P_{\max },\forall u \in {\cal U}, \label{const_Pwr} \\
& \sum\limits_{m = 1}^{M} {n_m^u}  + \sum\limits_{k = 1}^{K} \sum\limits_{i = 1}^I {o_{ki}^u} \le I \times {N_{RB}},\forall u \in \mathcal{U}, \label{const_RA}\\
& n_m^u \in \{ {n_{\rm{req}}},{n_{\rm{req}}} + 1, \ldots ,{N_{{\rm{RB}}}}\} , \forall m \in \mathcal {M}, u \in \mathcal{U}, \label{const_S}\\
& o_{ki}^{u} \in \{1,\dots, N_\mathrm{RB}\}, \forall k \in \mathcal{K}, i \in \mathcal{I}, u \in \mathcal{U}. \label{const_C}
\end{align}
\end{subequations}

For the $u$-th frame, (\ref{const_TP}) ensures each UE's throughput is no less than the throughput threshold $\gamma _k^u$, (\ref{const_Pwr}) is the power allocation for targets and UEs, and (\ref{const_RA}), (\ref{const_S}) and (\ref{const_C}) ensure the total allocated RBs and mini-slots do not exceed the maximum, where $n_{\rm{req}}$ means the minimum required RBs for each target. $n_{\rm{req}}$ is decided by the distance resolution $d_{\rm{res}}$ that the ISAC system wants to achieve, satisfying
\begin{equation}\label{SRBReq}
{n_{{\rm{req}}}} = \left\lceil {\frac{c}{{2{B_0}{d_{\rm{res}}}}}} \right\rceil.
\end{equation}
$c$ is the light speed. Problem ${\rm P}1$ is a mixed integer non-linear programming problem, which is non-convex considering the complex PCRB expression and the variable coupling. Besides, the dynamicity makes it more challenging to solve: the target movement and the time-varying communication channels.

%% file: 5Solution.tex
\section{Solution}
In this section, we develop a solution to the formulated problem ${\rm P}1$. The solution is supposed to decide how many resources are assigned to each target and UE for location tracking and communication at every frame, including the number of RBs, the number of mini-slots, the amount of power, and the selected beams for UEs. To decouple the problem from all the frames, we solve it for each frame as a sub-problem. In this case, the complexity is reduced with the price of obtaining the sub-optimum. For each sub-problem, we omit the index $u$. Considering the coupling of all the variables $\{ n_m,o_{ki},p_m,p_k,{\bf{f}}_k\}$, we separate them into two blocks $\{ n_m,o_{ki},p_k\}$ and $\{p_m,{\bf{f}}_k\}$, and apply BCD algorithm to solve each block iteratively while fixing the other variables. Details are introduced as follows.

\subsubsection{Sub-problem with respect to $\{n_m,o_{ki},p_k\}$}
Considering the complex matrix expression of PCRB in (\ref{P1Obj}), it is challenging to deal with it directly. Instead, we transform (\ref{P1Obj}) into a more easily handled algebraic form. Given fixed power $p_m$ for each target, the posterior FIM ${{\bf{J}}\left( {{\bf{\xi }}_m} \right)}$ can be rewritten as
\begin{equation}\label{FIMtransform}
{\bf{J}}\left( {{\bf{\xi }}_m} \right) = {{\bf{J}}_p}\left( {{\bf{\xi }}_m} \right) + {\left( {{\bf{\hat Q}}_m} \right)^{\rm{T}}}{\left( {{\bf{\hat \Sigma }}_m} \right)^{ - 1}}{\bf{\hat Q}}_m = {{\bf{E}}_m} + {\left( {n_m} \right)^2}{{\bf{V}}_m},
\end{equation}
where
\begin{equation}\label{Em}
{{\bf{E}}_m} = {{\bf{J}}_p}\left( {{\bf{\xi }}_m} \right) + {p_m}\left\| {{\bf{\hat H}}_m{\bf{f}}_m} \right\|_2^2/{\varpi _{nn}}{\bf{\bar q}}_2^{\rm{T}}{{{\bf{\bar q}}}_2},
\end{equation}
\begin{equation}\label{Vm}
{{\bf{V}}_m} = {p_m}\left\| {{\bf{\hat H}}_m{\bf{f}}_m} \right\|_2^2B_0^2{\bf{\bar q}}_1^{\rm{T}}{{{\bf{\bar q}}}_1}.
\end{equation}
${{{\bf{\bar q}}}_1}$ and ${{{\bf{\bar q}}}_2}$ are the first and second rows of ${{\bf{\hat Q}}_m}$. ${\rm{trace}}( {{\bf{J}}{{( {{{\bf{\xi }}_m}} )}^{ - 1}}} )$ can be further derived as~\cite{dong2022sensing}
\begin{equation}\label{EVconvex}
\begin{array}{*{20}{l}}
{{\rm{trace}}\left( {{{\left( {{{\bf{E}}_m} + {{\left( {{n_m}} \right)}^2}{{\bf{V}}_m}} \right)}^{ - 1}}} \right)}\\
{ = {\rm{trace}}\left( {{\bf{E}}_m^{ - \frac{1}{2}}{{\left( {{\bf{I}} + {{\left( {{n_m}} \right)}^2}{\bf{E}}_m^{ - \frac{1}{2}}{{\bf{V}}_m}{\bf{E}}_m^{ - \frac{1}{2}}} \right)}^{ - 1}}{\bf{E}}_m^{ - \frac{1}{2}}} \right)}\\
{ = {\rm{trace}}\left( {{\bf{E}}_m^{ - \frac{1}{2}}{\bf{G}}{{\left( {{\bf{I}} + {{\left( {{n_m}} \right)}^2}{\bf{\Lambda }}} \right)}^{ - 1}}{{\bf{G}}^{\rm{T}}}{\bf{E}}_m^{ - \frac{1}{2}}} \right)}\\
{ = {\rm{trace}}\left( {{{\bf{G}}^{\rm{T}}}{\bf{E}}_m^{ - \frac{1}{2}}{\bf{E}}_m^{ - \frac{1}{2}}{\bf{G}}{{\left( {{\bf{I}} + {{\left( {{n_m}} \right)}^2}{\bf{\Lambda }}} \right)}^{ - 1}}} \right)}\\
{ = \sum\limits_{j = 1}^4 {{{\left( {{{\bf{G}}^{\rm{T}}}{\bf{E}}_m^{ - \frac{1}{2}}{\bf{E}}_m^{ - \frac{1}{2}}{\bf{G}}} \right)}_{jj}}} {{\left( {1 + {{\left( {{n_m}} \right)}^2}{{\bf{\Lambda }}_{jj}}} \right)}^{ - 1}}}\\
{ = \sum\limits_{j = 1}^4 {{{\left( {{a_{mj}} + {b_{mj}}{{\left( {{n_m}} \right)}^2}} \right)}^{ - 1}}} }
\end{array}
\end{equation}
where ${\bf{E}}_m^{ - \frac{1}{2}}{{\bf{V}}_m}{\bf{E}}_m^{ - \frac{1}{2}} = {\bf{G\Lambda }}{{\bf{G}}^{\rm{T}}}$ is the eigendecomposition. $a_{mj} = {1 \mathord{\left/{\vphantom {1 {{{\left( {{{\bf{G}}^{\rm{T}}}{\bf{E}}_m^{ - \frac{1}{2}}{\bf{E}}_m^{ - \frac{1}{2}}{\bf{G}}} \right)}_{jj}}}}} \right.\kern-\nulldelimiterspace} {{{\left( {{{\bf{G}}^{\rm{T}}}{\bf{E}}_m^{ - \frac{1}{2}}{\bf{E}}_m^{ - \frac{1}{2}}{\bf{G}}} \right)}_{jj}}}}$ and $b_{mj} = a_{mj}{{\bf{\Lambda }}_{jj}}$. Therefore, the sub-problem with respect to $\{ n_m,o_{ki},p_k\}$ from the problem~${\rm P}1$ can be expressed as
\begin{subequations}\label{OptP11}
\begin{align}
& ({\rm P}1.1) \mathop {\min }\limits_{\{ n_m,o_{ki},p_k\} }  {\sum\limits_{m = 1}^{{M}} {\sum\limits_{j = 1}^4 {{{\left( {a_{mj} + b_{mj}{{\left( {n_m} \right)}^2}} \right)}^{ - 1}}} } } \label{P11Obj}\\
\mbox{s.t.}
&{ {\sum\limits_{i = 1}^I {(o_{ki}{B_0}){{\log }_2}\left( {1 + \frac{{{p_k}{{\left| {{\bf{h}}_{ck}{\bf{f}}_k} \right|}^2}}}{{o_{ki}{B_0}{N_0}}}} \right)} } } \ge \gamma _k, \forall k \in \mathcal{K},\label{const_TP11} \\
& \sum\limits_{m = 1}^{M} {p_m} + \sum\limits_{k = 1}^{K} {p_k} \le P_{\max }, \label{const_Pwr11} \\
& \sum\limits_{m = 1}^{M} {n_m}  + \sum\limits_{k = 1}^{K} \sum\limits_{i = 1}^I {o_{ki}} \le I \times {N_{RB}}, \label{const_RA11}\\
& n_m \in \{ {n_{\rm{req}}},{n_{\rm{req}}} + 1, \ldots ,{N_{{\rm{RB}}}}\} , \forall m \in \mathcal {M}, \label{const_S11}\\
& o_{ki}\in \{1,\dots, N_\mathrm{RB}\}, \forall k \in \mathcal{K}, i \in \mathcal{I} \label{const_C11}.
\end{align}
\end{subequations}
Then, we introduce a nonnegative auxiliary variable $z_{mj}$ such that ${( {a_{mj} + b_{mj}{{\left( {n_m} \right)}^2}})^{ - 1}} \le z_{mj}$. The problem~${\rm P}1.1$ becomes
\begin{subequations}\label{OptP12}
\begin{align}
({\rm P}1.2) & \mathop {\min }\limits_{\{ n_m,o_{ki},p_k,z_{mj}\} }  {\sum\limits_{m = 1}^{M} {\sum\limits_{j = 1}^4 z_{mj} } } \label{P12Obj}\\
\mbox{s.t.}
& \frac{1}{{z_{mj}}} - \left( {a_{mj} + b_{mj}{{\left( {n_m} \right)}^2}} \right) \le 0, \label{const_Z12}\\
& z_{mj} \ge 0, \label{const_zPos}\\
& (\ref{const_TP11}), (\ref{const_Pwr11}), (\ref{const_RA11}), (\ref{const_S11}), (\ref{const_C11}).
\end{align}
\end{subequations}
Note that the left-hand side of (\ref{const_Z12}) is a form of difference of convex (DC) function, to which the SCA technique can provide Karush-Kuhn-Tucker (KKT) conditions~\cite{bLuisTWC}. Therefore, we apply SCA to the problem~${\rm P}1.2$. In each iteration, ${{\left( {n_m} \right)}^2}$ is linearized by its first-order Taylor expansion as ${{{\left( {n_m^{*}} \right)}^2} + 2n_m^{*}\left( {n_m - n_m^{*}} \right)}$, where $n_m^{*}$ is the updated value of $n_m$ from the previous iteration. Therefore, the problem~${\rm P}1.2$ is further approximated in each iteration as:
\begin{subequations}\label{OptP13}
\begin{align}
& ({\rm P}1.3) \mathop {\min }\limits_{\{ n_m,o_{ki},p_k,z_{mj}\} }  {\sum\limits_{m = 1}^{M} {\sum\limits_{j = 1}^4 z_{mj} } } \label{P13Obj}\\
\mbox{s.t.}
& \frac{1}{{z_{mj}}} - \left( {a_{mj} + b_{mj}{{\left( {{{\left( {n_m^{*}} \right)}^2} + 2n_m^{*}\left( {n_m - n_m^{*}} \right)} \right)}}} \right) \le 0, \label{const_Z13}\\
& (\ref{const_zPos}), (\ref{const_TP11}), (\ref{const_Pwr11}), (\ref{const_RA11}), (\ref{const_S11}), (\ref{const_C11}).
\end{align}
\end{subequations}
The throughput constraint (\ref{const_TP11}) is the function of ${o_{ki}}$ and ${p_k}$, which we re-write as follows.
\begin{equation}\label{const_TP_trans}
f\left( {{o_{ki}},{p_k}} \right) = {\gamma _k} - \sum\limits_{i = 1}^I {({o_{ki}}{B_0}){{\log }_2}\left( {1 + \frac{{{p_k}{{\left| {{{\bf{h}}_{ck}}{\bf{f}}_k} \right|}^2}}}{{{o_{ki}}{B_0}{N_0}}}} \right)}  \le 0.
\end{equation}
It can be proved that the Hessian matrix of $f\left( {{o_{ki}},{p_k}} \right)$ with respect to ${o_{ki}}$ and ${p_k}$ is positive semi-definite. The detailed proof is omitted due to limited space. Thus, $f\left( {{o_{ki}},{p_k}} \right)$ and (\ref{const_TP_trans}) are convex. We apply the penalty method to deal with the constraint (\ref{const_TP_trans}) and move it to the objective function, changing the problem~${\rm P}1.3$ to the following version.
\begin{subequations}\label{OptP14}
\begin{align}
({\rm P}1.4) &\mathop {\min }\limits_{\{ {n_m},{o_{ki}},{p_k},{z_{mj}}\} } \sum\limits_{m = 1}^M {\sum\limits_{j = 1}^4 {{z_{mj}}} }  + \beta \sum\limits_{k = 1}^K {\max \left\{ {f\left( {{o_{ki}},{p_k}} \right),0} \right\}} \label{P14Obj}\\
\mbox{s.t.}
& (\ref{const_Z13}), (\ref{const_zPos}), (\ref{const_Pwr11}), (\ref{const_RA11}), (\ref{const_S11}), (\ref{const_C11}).
\end{align}
\end{subequations}
where $\beta  > 0$ is the penalty parameter, iteratively increasing to enforce the constraint (\ref{const_TP_trans}). Since the objective and constraints in problem~${\rm P}1.4$ are either convex or linear, the problem~${\rm P}1.4$ is convex. In each SCA iteration, the problem~${\rm P}1.4$ can be directly solved by CVX.

\subsubsection{Sub-problem with respect to $\{p_m,{\bf{f}}_k\}$}
Given the fixed $\{n_m,o_{ki},p_k\}$, the sub-problem with respect to $\{p_m,{\bf{f}}_k\}$ is decoupled. As introduced in the system model, the codebook-based beamforming is applied for ${\bf{f}}_k$. Denoting the codebook as $CB = {\{ {{{{\bf{\tilde f}}}_1},...,{{{\bf{\tilde f}}}_n},...,{{{\bf{\tilde f}}}_{{N_{CB}}}}} \}^{\rm{T}}}$, we perform a one-dimension search among $N_{CB}$ beamformer candidates and find the beamformer vector ${\bf{\hat f}}_k^{*}$ for each UE $k$ that maximizes the received signal-to-noise-ratio, namely
\begin{equation}
{\bf{\hat f}}_k^* = \mathop {\arg \max }\limits_{{{{\bf{\tilde f}}}_n} \in CB} \frac{{{p_k}{{\left| {{{\bf{h}}_{ck}}{{{\bf{\tilde f}}}_n}} \right|}^2}}}{{{o_{ki}}{B_0}{N_0}}}.
\end{equation}
Likewise, given the fixed $n_m$, the posterior FIM ${{\bf{J}}\left( {{\bf{\xi }}_m} \right)}$ can be rewritten as 
\begin{equation}\label{FIMtransform2}
{\bf{J}}\left( {{{\bf{\xi }}_m}} \right) = {{\bf{J}}_p}\left( {{{\bf{\xi }}_m}} \right) + {p_m}{{{\bf{\bar V}}}_m},
\end{equation}
where ${{{\bf{\bar V}}}_m} = {({{\bf{\hat Q}}_m})^{\rm{T}}}{({{\bf{\hat \Sigma }}_m})^{ - 1}}{{{{{\bf{\hat Q}}}_m}} \mathord{\left/ {\vphantom {{{{{\bf{\hat Q}}}_m}} {{p_m}}}} \right.\kern-\nulldelimiterspace} {{p_m}}}$. Note that (\ref{FIMtransform2}) has a similar form as (\ref{FIMtransform}). Therefore, a similar transformation as (\ref{EVconvex}) can be applied to the objective function. Then, the sub-problem with respect to $\{p_m,{\bf{f}}_k\}$ is given by
\begin{subequations}\label{OptP21}
\begin{align}
& ({\rm P}2.1) \mathop {\min }\limits_{\{ {p_m},{{\bf{f}}_k}\} } \sum\limits_{m = 1}^M {\sum\limits_{j = 1}^4 {{{\left( {{{\bar a}_{mj}} + {{\bar b}_{mj}}{p_m}} \right)}^{ - 1}}} } \label{P21Obj}\\
\mbox{s.t.}
& (\ref{const_TP11}), (\ref{const_Pwr11}).
\end{align}
\end{subequations}
where ${{{\bar a}_{mj}}}$ and ${{{\bar b}_{mj}}}$ are constants for $p_m$ with similar forms as in (\ref{EVconvex}). Detailed derivation is omitted due to limited space. After the beam is selected, the problem~${\rm P}2.1$ is convex concerning $p_m$, which can be solved by CVX. 

The overall algorithm of solving the problem~${\rm P}1.1$ is concluded as Algorithm~\ref{RAalgo}.
\begin{algorithm}[htbp]
\caption{BCD-based algorithm for solving problem ${\rm P}1.1$}
\label{RAalgo}
\begin{algorithmic}[1]
\STATE \textbf{Initialization}: Set $p_m$ and ${\bf{f}}_k$ with feasible values.
\REPEAT 
\STATE Solve problem ${\rm P}1.4$ by CVX iteratively until convergence to obtain $\{n_m,o_{ki},p_k\}$.
\STATE Solve problem ${\rm P}2.1$ by CVX to obtain $\{p_m,{\bf{f}}_k\}$.
\UNTIL the reduction of the objective function (\ref{OptP11}) is smaller than a threshold. \\		
\STATE \textbf{Output}: $\{n_m,o_{ki},p_k,p_m,{\bf{f}}_k\}$. \\
\end{algorithmic}
\end{algorithm}

%% file: 6Results.tex
\section{Performance Evaluation}
In this section, we evaluate our proposed solution via numerical simulations and compare it against the upper bound and the RFTEP benchmark. The upper bound is obtained with the maximum resources for each target and UE. The RFTEP benchmark means RBs and mini-slots are randomly allocated, the power is equally allocated, and the beam is selected by one-dimensional searching. The simulation parameters are set as Table~\ref{Tab1}. Additionally, the initial target locations are uniformly distributed within the range [0, 200m] along the X coordinate and [-100m, 100m] along the Y coordinate. The target velocities are uniformly generated between 0 and 30 km/h. The path between BS and targets is assumed to be a line-of-sight path. The path loss of UEs is modeled as $L(d) = {K_0}{d^{ - 2}}$, where ${K_0} = -30$ dB and $d$ is the distance between the BS and UE, uniformly distributed within the range [20 m, 40 m]. The penalty parameter $\beta$ increases five times at every iteration from 0.1. The convergence threshold is set as 0.001. The throughput threshold $\gamma _k$ is set as 1 Mbits if not specified. Next, we show the Monte Carlo simulation results with 100 iterations.

\begin{table}[htbp]
\caption{Simulation parameters}
\label{Tab1}
\setlength{\tabcolsep}{3pt} 
\centering
\begin{tabular}{cccc}
\hline
\textbf{parameter} & \textbf{value} & \textbf{parameter} & \textbf{value} \\ \hline
$N_t$ & 64 & $N_r$ & 64 \\
\rowcolor[HTML]{EFEFEF} 
carrier frequency & 39 GHz & sub-carrier space & 120 kHz \\
${\varpi _{nn}}$ & 4.14$^\circ$ & $P_{\max }$ & 53 dBm \\
\rowcolor[HTML]{EFEFEF} 
$K$ & 10  & $M$ & 10 \\
$I$ & 160~\cite{38211} & $N_\mathrm{RB}$ & 264 \\
\rowcolor[HTML]{EFEFEF} 
$N_0$ & $-$174 dBm/Hz & $N_P$ & 10 \\
$\rho_k$ & 0.75 & ${d_{\rm{res}}}$ & 1.5 m \\
\hline
\end{tabular}
\end{table}


\begin{figure*}[htbp]
    \begin{minipage}[t]{0.33\linewidth}
        \centering
        \includegraphics[width=1.02\textwidth]{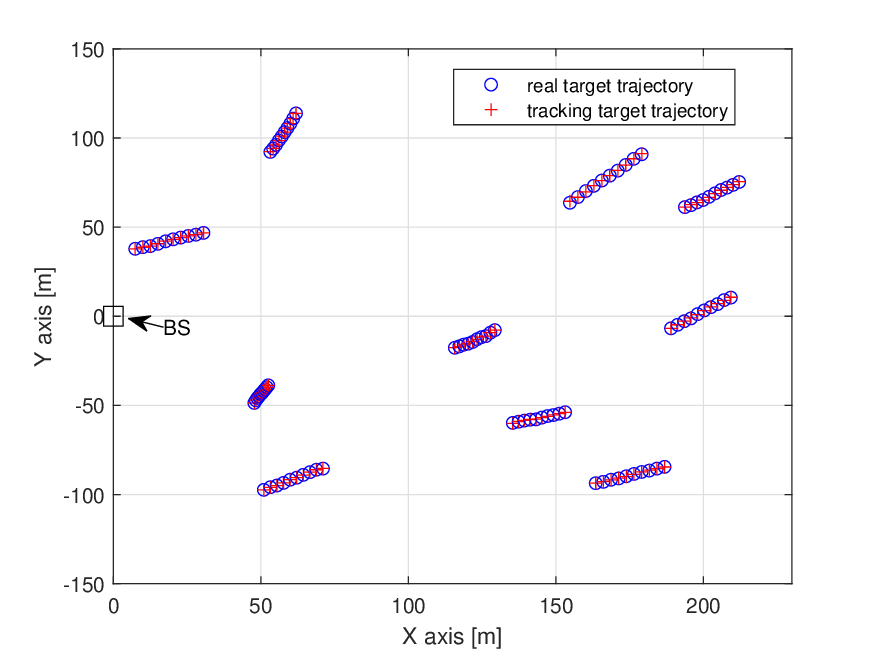}
        \caption{Location tracking results of targets.}
        \label{R_Trajectory}
    \end{minipage}
    \begin{minipage}[t]{0.33\linewidth}
        \centering
        \includegraphics[width=1.025\textwidth]{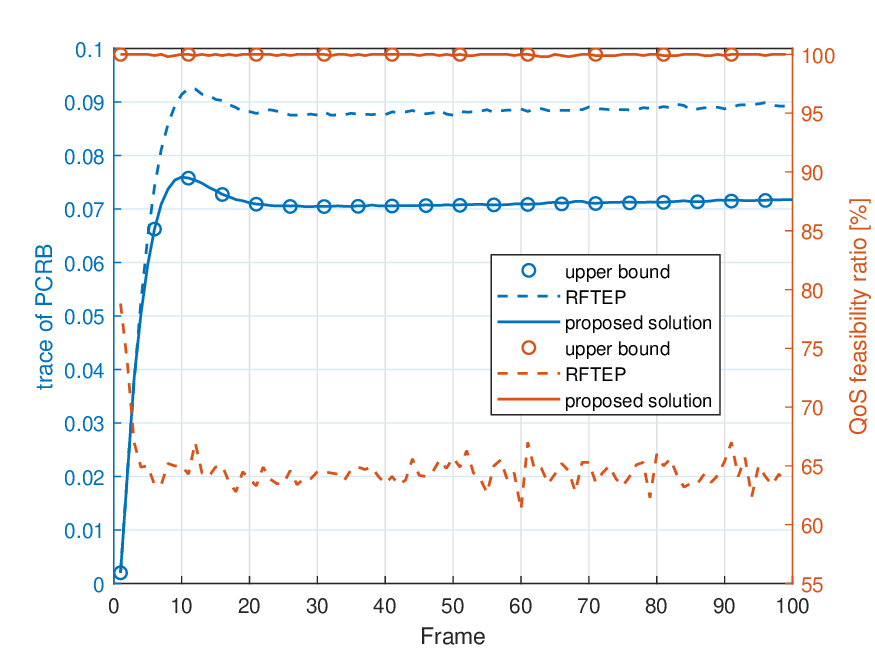}
        \caption{Performance comparison with benchmarks.}
        \label{R_COM}
    \end{minipage}
    \begin{minipage}[t]{0.33\linewidth}
        \centering
        \includegraphics[width=1.02\textwidth]{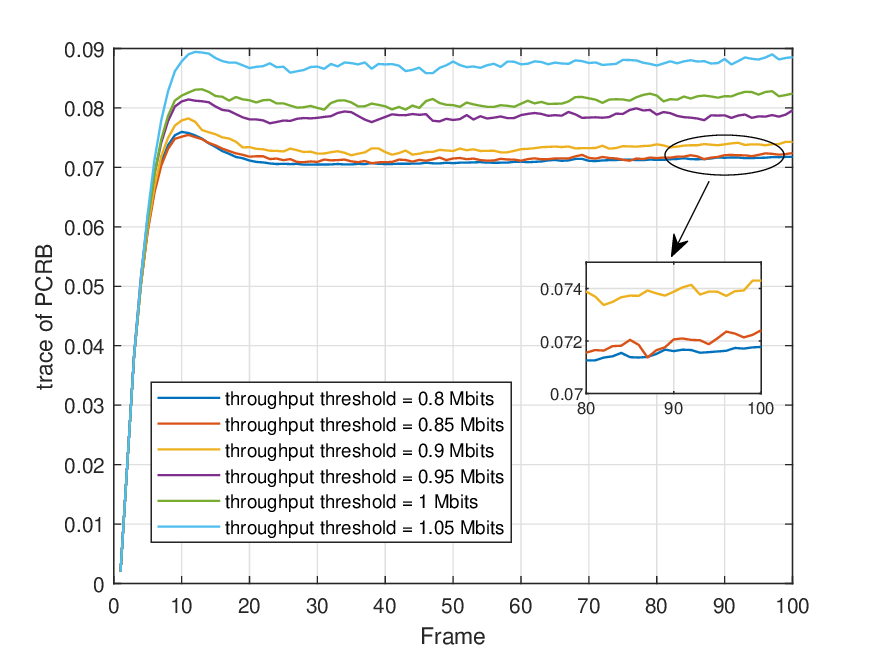}
        \caption{Trade-off of sensing and communication.}
        \label{R_tde_gamma}
    \end{minipage}    
\end{figure*}

Fig.~\ref{R_Trajectory} shows the location tracking results of 10 targets with different velocities using the proposed solution. It can be observed that all the targets are successfully tracked while satisfying the communication throughput requirements. Fig.~\ref{R_COM} presents the average sensing PCRB performance over all 10 targets and the average communication quality of service (QoS) feasibility ratio over all 10 UEs compared to the upper bound and RFTEP benchmarks. The QoS feasibility ratio indicates the percentage of UEs meeting the throughput constraints. The proposed solution reaches the upper bound at most frames in both sensing and communication. Compared to the RFTEP benchmark, the proposed solution achieves a 28.57\% performance gain in sensing and a 53.95\% gain in communication. In Fig.~\ref{R_tde_gamma}, the sensing PCRB performance worsens as the communication throughput threshold increases from 0.8 Mbits to 1.05 Mbits. This is because more resources are allocated to communication UEs to satisfy the increasing throughput constraint, leaving fewer resources for optimizing the sensing PCRB. This phenomenon illustrates the trade-off between sensing and communication in the ISAC system. In summary, simulation results verify the effectiveness of the proposed solution.

%% file: 8Conclusion.tex
\section{Conclusion}
In this work, we studied resource allocation in a communication and target tracking scenario, considering the movement of targets and the time-variant communication channel. To ensure the performance of both sensing and communication, we minimized the trace of PCRB for all targets subject to the throughput threshold and resource allocation constraints, which is non-convex. We proposed a BCD algorithm based on the penalty method, SCA technique, and one-dimension search. Simulation results demonstrated the validity of the proposed solution and the trade-off between sensing and communication.